\documentclass[12pt]{article}
\usepackage{authblk}
\usepackage{amssymb, amsmath,amsfonts, bm, eurosym,geometry,ulem,graphicx,caption,mathtools,natbib, xcolor,setspace,comment,pdflscape,array, parskip, bm, csquotes, multirow, booktabs, bbm, nicefrac}

\usepackage[]{footmisc}
\usepackage[colorlinks=true]{hyperref}
\usepackage{footnotebackref}
\usepackage{dirtytalk}
\usepackage[]{threeparttable}

\usepackage[capposition=bottom]{floatrow}

\usepackage[toc,page]{appendix}

\usepackage{color, amsthm}
\usepackage{tikz, subfig, cleveref}
\usetikzlibrary{positioning, patterns, decorations.pathreplacing, calligraphy, arrows.meta, bending, decorations.markings}
\tikzset{
    arc arrow/.style args={%
    to pos #1 with length #2 and options #3}{
    decoration={
        markings,
         mark=at position 0 with {\pgfextra{%
         \pgfmathsetmacro{\tmpArrowTime}{#2/(\pgfdecoratedpathlength)}
         \xdef\tmpArrowTime{\tmpArrowTime}}},
        mark=at position {#1-\tmpArrowTime} with {\coordinate(@1);},
        mark=at position {#1-2*\tmpArrowTime/3} with {\coordinate(@2);},
        mark=at position {#1-\tmpArrowTime/3} with {\coordinate(@3);},
        mark=at position {#1} with {\coordinate(@4);
        \draw[-{Stealth[length=#2,bend,#3]}]
        (@1) .. controls (@2) and (@3) .. (@4);},
        },
     postaction=decorate,
     }
}
\tikzstyle{line}=[draw]
\usepackage{pgfplots, tablefootnote}
\usepgfplotslibrary{fillbetween}
\usepgflibrary{arrows}
\usepgflibrary{decorations.pathmorphing}
\usepgflibrary{shapes.geometric,shapes.misc}
\pgfplotsset{compat=1.3}

\graphicspath{ {./images/} }

\newtheorem{corollary}{Corollary}
\newtheorem{proposition}{Proposition}

\newcolumntype{L}[1]{>{\raggedright\let\newline\\arraybackslash\hspace{0pt}}m{#1}}
\newcolumntype{C}[1]{>{\centering\let\newline\\arraybackslash\hspace{0pt}}m{#1}}
\newcolumntype{R}[1]{>{\raggedleft\let\newline\\arraybackslash\hspace{0pt}}m{#1}}

  {\list{}{\leftmargin=0.3in\rightmargin=0in}\item[]}%
  {\endlist}

\hypersetup{
    pdftoolbar=true,        
    pdfmenubar=true,        
    pdffitwindow=false,     
    pdfstartview={FitH},    
    pdftitle={title},    
    pdfauthor={Salvatore Mazzarino},     
    pdfsubject={Subject},   
    pdfcreator={Salvatore Mazzarino},   
    pdfproducer={Salvatore Mazzarino}, 
    pdfkeywords={Green Networking} {Mobile Cloud} {Network Coding} {Energy}, 
    pdfnewwindow=true,      
    colorlinks=true,       
    linkcolor=blue,          
    citecolor=blue,        
    filecolor=blue,      
    urlcolor=blue           
}

\makeatletter
\renewcommand\hyper@natlinkbreak[2]{#1}
\makeatother

\begin{document}

\title{Criminal Property Rights Suppress Violence in Urban Drug Markets: Theory and Evidence from Merseyside, U.K.}

\author[a]{Paolo Campana}
\author[a,b]{Andrea Giovannetti}
\author[c,d]{Paolo Pin}
\author[c]{Roberto Rozzi}
\affil[a]{University of Cambridge, UK}
\affil[b]{Australian Catholic University, Australia}
\affil[c]{Dipartimento di Economia Politica e Statistica, Universit\`a di Siena, Italy}
\affil[d]{BIDSA,  Universit\`a  Bocconi, Milan, Italy}
\date{\today}

\maketitle

\begin{abstract}
    {In this work, we provide empirical evidence on organized criminal groups' (OCGs) behavior across the Liverpool area in the U.K. (Merseyside). We find that violent crimes concerning OCGs concentrate in the areas yielding the highest revenue, while OGCs primarily control areas yielding middle or low revenue. We explain and generalize these empirical observations with a theoretical model examining how OCGs strategically select which area to exploit based on expected revenue and the presence of other OCGs. We prove our results for three OCGs analytically and extend them to larger numbers of OCGs through numerical simulations. Both approaches suggest that, when the frequency of OCG activity is sufficiently high, each OCG controls one area, while the violence between OCGs remains low across all areas. When the frequency of OCG activity reduces, violent collisions between OCGs occur in the areas yielding the highest revenue, while some OCGs retain control over the medium-revenue areas. Our results suggest important policy recommendations. Firstly, if interventions are only violence-driven, they might miss critical underlying factors. Secondly, police operations might have unintended negative externalities in other areas of a city when they target criminal property rights, like increased violence in the areas yielding the highest revenue.}
    \noindent \\
\noindent \textsc{Keywords:Criminal property rights $|$ Shadow institutions $|$ Territorial control $|$
Territorial violence $|$
Dynamic game theory modeling}
\end{abstract}

\section{Introduction}

Property rights are a key feature of modern societies, and with them the institutional infrastructures of governance developed over centuries \citep{williamson1996mechanisms,dixit2004lawlessness}. Such infrastructures are often based on the presence of a legal system that defines and protects property rights, thus ensuring exchanges and peaceful interactions \citep{dixit2004lawlessness}. However, not all activities take place within the remit of the law. Illegal markets are a chief example. While such activities are not negligible for their size (revenues) and adverse consequences, their underlying property rights' processes have received very limited formal explorations. Yet, we know that the absence of institutions, such as a legal system, does not prevent individuals or organizations from finding ways to exert control over exploitable resources, for example through the formation of informal property rights \citep{hirshleifer1995anarchy, hafer2006origins}. Understanding the dynamics of the formation of property rights in the absence of institutions provides key insights into the workings of societies, including their size of activities that place themselves outside the realm of the law \citep{badillo2023governing,blattman2024gang}.

In this paper, we study the behavior of organized crime groups (OCGs) operating in contexts where resource possession cannot be established or enforced by any (legal) authority but  must instead be self-regulated \citep{reuter1983disorganized, reuter1985organization, schelling1971business, campana2013cooperation}. In line with \cite{schelling1971business}, \cite{reuter1983disorganized}, \cite{gambetta1993sicilian}, \cite{varese2001russian} and \cite{campana2011eavesdropping}, and the widely agreed international definition by the United Nations \citep{UNTOC2000}, we interpret OCGs as (rational) revenue-seeking entities operating outside the realm of the law. In this work, we analyze empirical evidence on behavior of organized criminal groups in Merseyside, U.K., through the lenses of a theory of dynamic territory occupation.

In our empirical analysis of OCG activity across Merseyside, we used three key indicators for each territory: the number of OCGs operating in that area, the level of OCG-related violence, and the average control exerted by a single OCG in that area. We call the latter measure \textit{streak}, and we define a streak as the number of consecutive days in which OCG-related crimes in the area are committed exclusively by members of a single OCG (note that days without recorded OCG activity do not interrupt a streak; only a crime attributed to a different OCG does -- see Appendix for details on this metric). We focus on property right processes in one specific illegal market, i.e., the drug markets (the largest and most profitable according to many scholars: \cite{reuter2014drug}). We rank territories based on the number of drug trafficking crimes reported by Merseyside Police, and take drug trafficking levels as a measure of a territory's revenue.

We find that while OCGs concentrate in and exert high levels of violence in the most remunerative drug trafficking areas, they fail to establish property rights over these territories.\footnote{Our empirical findings are descriptive in nature. While they document striking regularities in the spatial and strategic behavior of OCGs, they do not establish causal relationships between drug revenue, violence, and territorial control.}
Instead, territorial control is achieved primarily in medium- to low-revenue areas. This pattern is reflected in crime streaks, which are longer in medium- to low-revenue areas than in the high-revenue ones. The latter are also the areas in which violence is lower. These findings suggest that the most peaceful territories are subject to OCGs control, pointing to a violence-suppressing effect of criminal governance.

This static picture hides the presence of complex, longitudinal, spatial dynamics. OCGs are not confined to a single territory, but they can actively seek to expand their operations looking for further opportunities and financial rewards. However, in their simultaneous quest for a revenue-driven expansion of their criminal activities and for establishing property rights on markets and territories \citep{schelling1971business, campana2018organized}, OCGs can get entangled in violent dynamics. We further explore such dynamics through a model of OCG behavior premised on the idea that OCGs, as revenue-driven organizations, occasionally move out of their turf to run operations in different areas of a city. We distinguish between three types of areas: high, middle and low revenue areas. If two OCGs arrive in the same area, the last arrived incurs the 
cost of shifting and leaves \citep{kirchmaier2024commuting}. OCGs act opportunistically, exploring areas based on revenue potential and their belief that the area is unoccupied. In our model, OCGs establish property rights through de facto norms: an OCG does not try to exploit a territory if they expect another OCG to exploit it with a sufficiently large probability.

The main insight of our theoretical model indicates that as the frequency of activity of OCGs reduces, some OCGs establish property rights over the middle revenue areas while others have occasional violent confrontations for the control of the high revenue areas. This result has two important policy implications. Firstly, violent dynamics change depending on the extent to which OCGs are able to exert property rights over a territorially-based drug market (and possibly illegal markets in general). Middle revenue areas might not be a primary target of enforcement as the level of violence is lower; however, they can still possess a high level of criminal activity and, worse still, effective property rights over illegal markets. If interventions are primarily violence-driven, they might miss very critical situations of illegal governance by OCGs \citep{CampanaMeneghiniKnisley2025}. Secondly, police operations reducing OCGs' activity might induce a negative externality, e.g., increased violence in high revenue areas. Taken together, this shows that interventions need to be carefully thought out and to take into consideration effects beyond the single OCG targeted.

Our work builds on the long-standing economic literature on the appropriation of goods in the absence of formal institutions to enforce possession. Early works in this field examined economies where property rights are not initially defined but can be enforced through the allocation of scarce resources \citep{hirshleifer1995anarchy,grossman1995swords}. More recent studies have incorporated time into these models, exploring the evolution of property rights over time \citep{maxwell2005continuing,eggert2011dynamic, sekeris2024conflict}. While various works have shown that anarchic competition can lead to the emergence of property rights as a solution to conflicts over goods  \citep{skaperdas1992cooperation,grossman2001creation}, there remains debate on whether such allocations are efficient. On the one hand, efficiency may be achieved when agents have sufficient information \citep{piccione2007equilibrium, schwarz2019jungle}, but on the other hand, efficiency might not be granted, in the presence of asymmetric information \citep{hafer2006origins, adamson2024territory, deng2024contests}.

Since, in our model, agents (i.e., OCGs) cannot defend a resource (i.e., an area) once they have exploited it, the competition happens in a so-called state of amorphy rather than in one of anarchy \citep{hirshleifer1995anarchy}. In our model, property rights emerge as \textit{de facto} norms between OCGs that might not try to exploit an area of the city if other OCGs repeatedly exploit such area through time. This approach better reflects the dynamics of OCGs' conflict over city areas \citep{felson2006ecosystem, zimring2020crime}. Differently from anarchic systems \citep{hirshleifer1995anarchy,hafer2006origins}, amorphy results in agents not always being able to establish property rights over all resources.

The narrative of our paper parallels evolutionary models across different disciplines such as biology \citep{smith1974theory}, sociology \citep{foley2018conflict,foley2021avoiding}, anthropology \citep{bowles2013coevolution}, or economics \citep{lipnowski1983voluntary, kimbrough2020war}. We study the conditions under which agents violently collide or rather establish property rights over resources. Finally, some papers in the literature on dynamic models have studied the problem of agents appropriating or exploiting different kinds of resources \citep{hanaki2011born, chakraborti2015statistical}. We study a dynamic model where agents only recall the most recent period in which they attempted to exploit a resource and this information is sufficient in equilibrium. By doing so, agents that exploit the high-revenue resource might differ depending on the activity level of the agents themselves.

\cite{schelling1971business} defined the business of organized crime as the business of control over criminal markets. In our paper, we show that, while OCGs do aspire and, in some cases, succeed in establishing control over a spatially-situated resource, they are limits to establishing such control. Rather than focusing just on the spatial diffusion of crime through time \citep[][]{mohler2011self,short2010dissipation, brantingham2012ecology}, which is what much of the ecological criminology literature does, our approach has the advantage of being able to capture the \textit{incentives} behind OCGs' decision-making process. Additionally, we add a temporal dimension to our analysis by relying on a dynamic model of OCG behavior. Our findings highlight the complex and evolving relationship between places, OCGs' activities, including their quest for establishing property rights over resources, i.e. drug markets, and violence. They also show the complexity of policy interventions, with poorly designed policies potentially leading to unintended urban instability and intensified violence in high-revenue areas.

The remainder of the paper is organized as follows. Section~\ref{sec:mem} documents empirical regularities in the behavior of OCGs in Merseyside, which serve as a motivation for our general theory of property rights developed in Section~\ref{sec:mod}.  In Subsection~\ref{subsec:mod}, we analytically characterize the equilibrium for three OCGs, and in Subsection~\ref{subsec:sim}, we extend our results to larger numbers of OCGs through numerical simulations. Lastly, we discuss policy implications in Section~\ref{sec:disc}.

\section{Empirical Motivation}\label{sec:mem}

In this section, we document empirical regularities in the behavior of OCGs across Merseyside. These patterns motivate our general theory of property rights, presented in Section~\ref{sec:mod}, which formalizes and extends the mechanisms suggested by the data.

\textbf{Data.} We illustrate the complex relationship between organized crime and places by leveraging a granular dataset collecting all crime activities recorded in a large area of the United Kingdom, the Merseyside county. Merseyside is the fourth most populated metropolitan county in the U.K. ($1.38$ million population; surface of 645km$^2$, with Liverpool being the largest city. Importantly for our work, Merseyside features the highest number of OCGs per million people in England and Wales \cite{her2017peel}, $N=134$ groups. For comparison, this rate is twice as high as the national average and a quarter higher than Greater London. The unit of analysis for this study is a cross-section of all crime activities recorded in the $M=180$ areas of Merseyside between 1/1/2015 and 4/30/2018.

Our notion of area coincides with the \textit{Middle Layer Super Output Areas} (MSOAs), demographically stable small-area census units containing around $8,000$ inhabitants each. We characterize MSOAs through a novel dataset based on the complete set of crime reports involving at least one identified offender handled by Merseyside police (MP). This is the most comprehensive source of crime data available for Merseyside\footnote{MP is the primary source of Merseyside data for commercial and non-commercial crime data providers (e.g. \textit{Databank of U.K. Police}, \textit{Economic Policy Centre}).}.
The dataset contains $106,610$ incidents in which at least one individual (out of a population of $K = 62,948$ offenders) has been arrested, cautioned, charged as well as interviewed or suspected. Each incident is geo-tagged and linked to one offense category from the taxonomy of 384 items of English criminal law. Each offender $i$ linked to the incident is characterized with two critical pieces of information: a personal reference number and, in case the individual has been identified as an OCG member by the MP analysts, the unique OCG reference number.
The dataset has two limitations: first, it does not contain information about victims, if any. Second, data storage follows a first-in, first-out architecture with only one item per feature admitted per time. As a result, in our data, any offender can be associated with at most one OCG.

\begin{figure}[t!]
\centering
\includegraphics[width=1.15\linewidth]{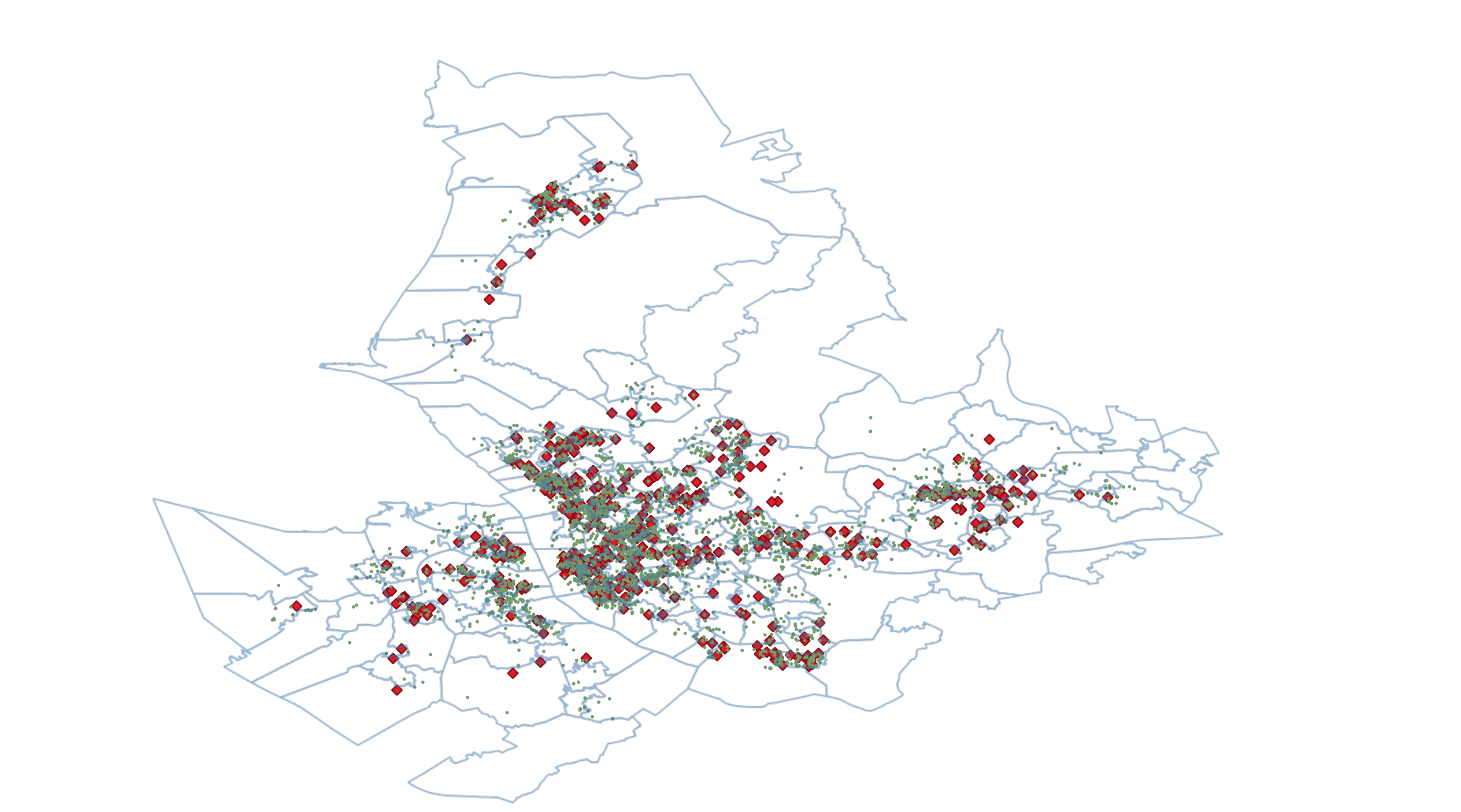}
\caption{Drug offenses and OCG-related violence in Merseyside. Each area marked by a blue contour represents one of the Middle Layer Super Output Areas (MSOAs) associated with Merseyside metropolitan county in the ONS coding system. Green dots identify a drug-related offence, whereas red squares capture a violent offence involving at least one individual associated with an OCG.
\label{fig:map_mersey} }

\end{figure}

Relevant to our study, the dataset contains 22,322 drug-related\footnote{In the U.K. system, controlled substances are categorised in Class A and Class B drugs. Class A refers to hard drugs such as crack cocaine, cocaine, MDMA, heroin, LSD, mushrooms, methadone and methamphetamine (crystal meth). In contrast, Class B refers to soft drugs such as Amphetamines, barbiturates, natural and synthetic cannabis, cathinones, as well as codeine, ketamine, anabolic steroids, and benzodiazepines.} incidents and 626 cases of serious violence\footnote{This corresponds to offences associated to the class \textit{Violence With Injury}, particularly, \textit{Wounding with Intent to do Grievous Bodily Harm} ($36.3\%$), \textit{Assault Occasioning Actual Body Harm} ($34.9\%$) and \textit{Murder} or \textit{Attempted Murder} ($17.3\%$).} where at least one of the offenders is a member of an OCG. In Figure \ref{fig:map_mersey}, OCG-related violence and drug offences, respectively given by green dots and red squares, have been superimposed against the MSOA map of Merseyside (in blue). Geographically, the relationship is complex. While some clustering of the two classes of crime is apparent in the central MSOAs of Merseyside, this does not appear to be a general phenomenon, as several areas characterised by dense drug activities show a limited amount of violence.


\begin{figure}[t!]
\centering
\includegraphics[width=0.9\linewidth]{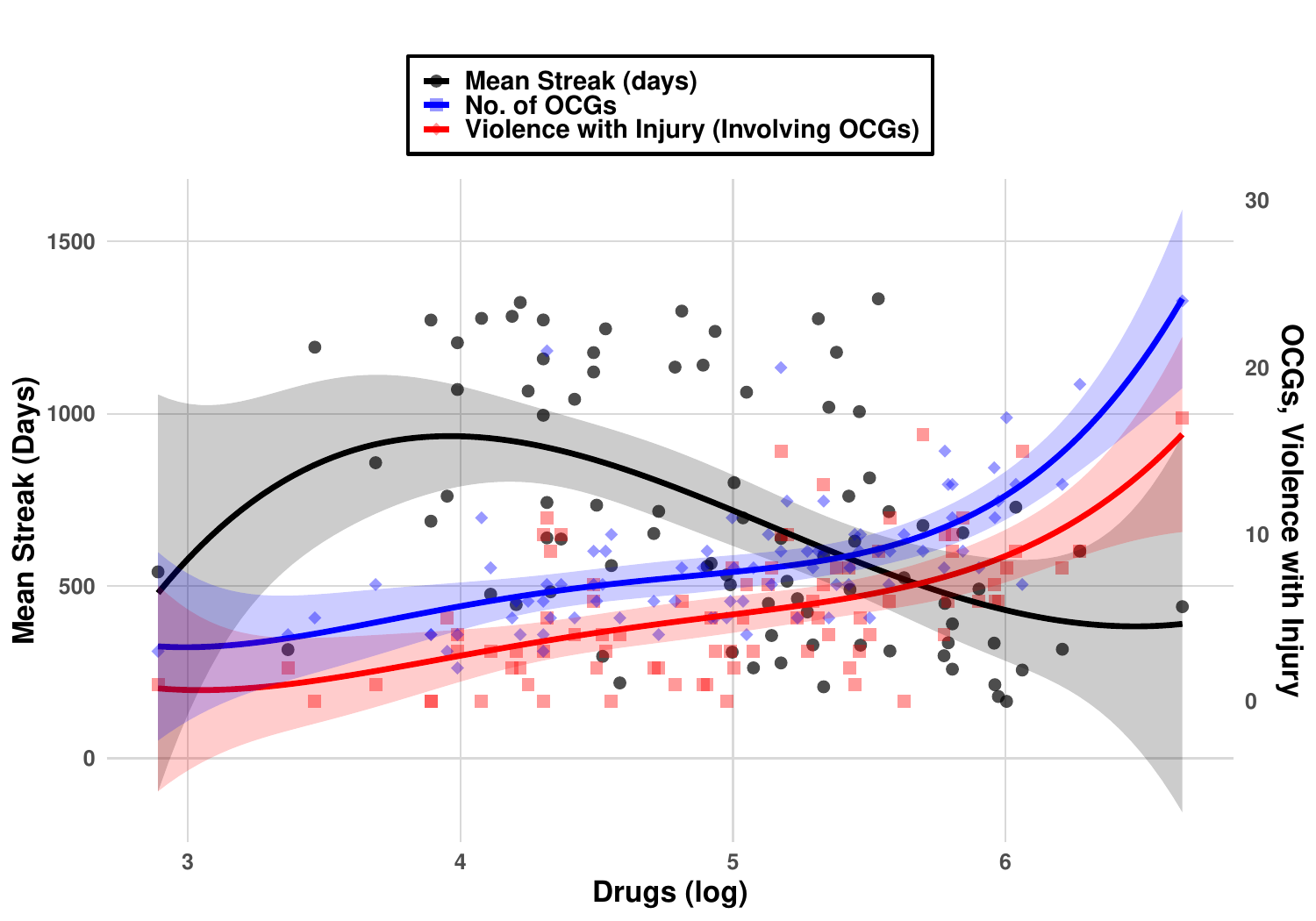}
\caption{In this figure, the x-axis is given by the log of the total count of drug dealing events recorded in each area of Merseyside.\textit{(Left Axis.)} Scatterplot of the episodes of OCG-induced serious violence and polynomial fit. \textit{(Right Axis.)} Scatterplots of the average number of OCGs and average number of areas per OCG (respectively plotted in blue and black) and their corresponding polynomial fit and confidence interval. Each area corresponds to a specific Middle Layer Super Output Area (MSOA) in the ONS coding system. Polynomial fits are obtained with polynomials of degree four.
\label{fig:dual_axis_plot} }
\end{figure}

\noindent \textbf{Methods.} We represent OCGs through the sum of crime activities carried out by their members and map these quantities into geographic areas. Formally, given any OCG $n \in N$ and area $m \in M$, define  $C^\ell_{m}$ as the count of crimes of type $\ell \in L$ committed in area $m$. Similarly, let $C^{\ell,ocg}_{m,n}$ and $C^{\ell,ocg}_{m}$ be\footnote{Formally, given any offender $k \in K$,  $C^\ell_{m} \equiv \sum_{k=1}^{K} C^\ell_{k,m} $  and $C^{\ell,ocg}_{m,n}   = \sum_{k=1}^{K} \left( g_{k,n} C^\ell_{k,m} \right)$, where $g_{k,n}$  takes value of $1$ if individual $k$ has been associated to OCG $n$ and $0$ otherwise. Last, define  $C^{\ell,ocg}_{m} \equiv \sum_{n=1}^{N} C^{\ell,ocg}_{m,n}$.} the count of crimes of type $\ell$ committed by associates of OCG $n$ in area $m$ and the sum of all OCG-related crimes of type $\ell$ committed in area $m$, respectively. In the following, we will specialize $\ell$ alternatively to crimes of serious violence, $\ell = \text{V}$ or drugs, $\ell =\text{D}$.

Then, given $\phi(x)$ a simple indicator function taking value $1$ for $x>0$ and $0$ otherwise, we define
\begin{equation} 
N_m  = \sum_{\ell=1}^L \sum_{n=1}^{N} \phi\left( C^{\ell, ocg}_{m,n} \right)
\end{equation}
 as the total number of OCGs operating in an area $m$.

For each area of Merseyside, we track the behavior of OCGs along three dimensions: $(i)$ a simple count of the number of OCGs, mirroring the concentration of OCGs operating in that area $(ii)$ the amount of recorded OCG-related violence; $(iii)$ the average length of a \textit{streak}, that is the number of days that occur between the arrest of individuals associated to \textit{different} OCGs, our measure of area control\footnote{Details about the computation of this metrics, statistics and a worked-out example are reported in the Appendix.}. To explore the opportunistic nature of OCGs, we then rank areas according to a simple measure of \textit{revenue}: the number of crimes recorded by MP on its territory involving drug dealing.

Figure~\ref{fig:dual_axis_plot} collects three scatterplots, each mapping one of the three indicators defined above against the log of the number of drug-related crimes. For each data cloud, we fit\footnote{In untabled analysis, we show that similar results, in qualitative terms, are obtained with degree 3 and 5.} a polynomial model of degree 4. Each polynomial fit (given by the solid lines), as well as the associated confidence intervals (given by the shadowed areas), are superimposed onto the scatterplots. Indicators $(i)$, $(ii)$, and $(iii)$, and the associated polynomial fit and confidence intervals are drawn in blue, red, and black color, respectively.

We make two observations. First, we note that the concentration of OCGs and the amount of OCG-related violence correlate positively, with both concentration and violence increasing in the value of areas. While the increasing concentration is possibly hinting at the opportunistic nature of OCG decision-making (with more OCGs being active in high-revenue areas), the higher levels of violence consistently observed in areas characterized by higher drug trafficking point to the higher competitive pressure faced by OCGs in these areas. Second, the streak indicator is higher for areas with medium rather than high revenue. Taken together, these observations loosely identify three blocks of areas: a \textit{low revenue} block (roughly corresponding to areas featuring up to $C^{D,ocg}_{m} = 54$ cases of drug dealing), where a sparser presence of OCGs and relatively low turnaround between OCGs mirror scarcer economic opportunities; a \textit{medium revenue} block (corresponding to areas where $54 < C^{D,ocg}_{m} \leq 155$), populated by a stable number of OCGs enjoying low violence and long streaks; and a \textit{high revenue} segment (corresponding to areas where $C^{D,ocg}_{m} > 155$) in which violence and OCG concentration are increasing in the area's revenue and turnovers are shorter.

\section{A General Theory of Property Rights}\label{sec:mod}

Empirical evidence from Merseyside reveals complex OCG behaviors, raising questions about turf selection and conflict. We show that these patterns can emerge as equilibrium outcomes of a game-theoretic model with endogenous property rights. As a working definition, in the remainder of the paper, we will hold that  \emph{property rights} are established over an area if, in equilibrium, only one OCG seeks to exploit such an area. Conversely, property rights are absent in all those areas where, in equilibrium, exploitation is contended between multiple OCGs. We provide the statement and intuition of our theoretical results in the main text, leaving all the formal proofs in the supplementary materials.

\subsection{The model with 3 OCGs and 3 areas}\label{subsec:mod}

We start by considering a stylized city made of $M=3$ areas and $N=3$ OCGs, denoted $1,2, \text{ and } 3$. Each OCG is based in a turf located outside the city. At any given time, each OCG can be either in its turf or in the city. Upon entering the city, an OCG will \textit{sequentially} explore the various areas until a suitable one is found. Subsequently, it will \textit{occupy} that area (e.g., engaging in drug dealing or other illicit activities). Time is continuous: each OCG leaves its turf at an exponential rate $\eta$ and goes back to it at a fixed exponential rate of $\gamma$, which we fix to $\gamma=1$ for convenience. Revenue is heterogeneous across areas and is captured by parameters $u_1$, $u_2$, and $u_3$, for area $1$, $2$, and $3$, respectively. In any period, an OCG cannot exploit an area that another OCG has already occupied in that period: the last to arrive incurs a cost $c>0$ (for example, a \emph{collision} cost, as we hold below) and leaves the area empty-handed. We assume that $u_1 > u_2 > u_3 > c$.

OCGs only know what they see in each area they visit at the time they visit it. Formally, let $z_i^m$ be the state of the area $m^*$ last exploited by OCG $i$ and let period $t_i^m$ indicate when area $m$ was last visited by $i$. Then, at any point in time, the information set of OCGs consists of {\em i)} the time $t_i^m$ when area $m$ was last seen and {\em ii)} the state $z_i^m$ in which it was left at that time. For example, for the area $m^*$ last exploited by OCG $i$, $z_i^{m^*}=0$ and $t_i^{m^*}$ is the time when OCG $i$ left it. For an area $m$ which was found occupied in a search attempt at time $t_i^m$, we set $z_i^m=1$. Last, let $p^m$ be the stationary state probability that area $m$ is unoccupied. We assume that $p^m$ is common knowledge (i.e., all OCGs will be able to learn it).

We study OCG decision-making by looking at the strategies pursued by OCGs in equilibrium. For any OCG $i$, a \textit{strategy} is a rule that, given the OCG's information set, prescribes the sequence of exploration of the three areas (i.e. it assigns a permutation of the three areas). 

\paragraph{OCGs' Exploration Patterns.} Since $p^m$ is common knowledge, the belief that OCG $i$ holds on area $m$ being unoccupied at time $t$ given the information that OCG $i$ has at this time is encapsulated in the following probability

\begin{equation}
\label{qim}
q_i^m(t)=p^m \left( 1 - e^{ - \frac{t-t_i^m}{1-p^m}} \right)+\left(1-z_i^m\right)e^{ - \frac{t-t_i^m}{1-p^m}}.
\end{equation}

Given these pieces of information, OCGs play their strategy.

By solving the game, we identify the optimal exploration strategy devised by any OCG $i$, given the information encoded in her beliefs $q_i^m$. Through the optimal search strategy, each OCG $i$ ranks areas by their \textit{expected return} $u_m - c/q^m_i$, in decreasing order. OCGs prioritize areas that provide a higher revenue against the expected costs of finding that area occupied. If the OCG finds an area occupied, such OCG will downgrade the area's expected return for the forthcoming round of explorations and, at the same time, will move on to exploring a less valuable area (see Lemma D.1 in supplementary material). 
Importantly, while the expected exploration strategy is the same for all OCGs, exploration patterns will likely differ across OCGs due to the heterogeneity in belief formation.

\paragraph{Equilibrium Analysis.} Having identified the mechanism of exploration used by OCGs in equilibrium, we introduce the core of our analysis, which consist of exploring the nexus between $\eta$ and three key variables in the formation of \textit{de facto} property rights over areas: concentration and violence levels, and the mean streak over an area.

In the supplementary materials, we derive results about the optimal exploration pattern of OCGs as a function of $\eta$ (see Lemma D.1 to D.3). If $\eta > \frac{u_1 - u_3}{c}$ OCGs ‘specialize’ into different areas, focusing on the area they can control without facing competition. It follows that the more active OCGs are, the more likely they are to establish property rights across the city. For $\max\{\frac{u_1 - u_2}{c}; \frac{u_2 - u_3}{c}\} < \eta < \frac{u_1 - u_3}{c}$, criminal groups do not establish property rights over the best area, but over the middle one only. Lastly, if $\eta < \max\{\frac{u_1-u_2}{c}; \; \frac{u_2-u_3}{c}\}$, the frequency of activity of OCGs is so small that each OCG can find the right belief to start from area $1$, hence, no property rights are established over areas. For simplicity let us call $\underline{\eta}=\max\{\frac{u_1-u_2}{c}; \; \frac{u_2-u_3}{c}\}$ and $\overline{\eta}=\frac{u_1 - u_3}{c}$.


\paragraph{Concentration Levels.} Let us call $O_m(\eta)$ a function that gives for each value of $\eta$ the concentration level of area $m$.
\begin{proposition}\label{prop_3ag_occ}

$\text{ }$


\begin{itemize}
    \item If $\eta > \overline{\eta}$, $O_1(\eta)=O_2(\eta)=O_3(\eta)$;
    \item If $\eta < \overline{\eta}$, $ O_3(\eta) < O_2(\eta) < O_1(\eta)$.
\end{itemize}

\end{proposition}

\begin{figure}[ht!]
    \centering
    \begin{tikzpicture}[dot/.style={circle,draw=black, fill,inner sep=1pt},scale=1.3]
    \draw[thick, ->] (0,0) -- (0,5);
    \node[thick] at (-0.3,4.8) {$\mathbb{R}$};
    \node[dot, red] at (4.3,5) {};
    \node[red] at (6,5) {Freq. of violence};
    \node[dot, blue] at (4.3,4) {};
    \node[blue] at (6.1,4) {Concentration lev.};
    \node[dot] at (4.3,3) {};
    \node[] at (5.6,3) {Mean streak};
    \draw[thick, ->] (0,0) -- (5.1,0);
    \node[align=left] at (5,-0.4) {$u_m$};
    \draw[thick] (1.2,-0.1) -- (1.2,0.1);
    \node[align=left] at (1.2,-0.3) {$u_3$};
    \draw[thick] (2.4,-0.1) -- (2.4,0.1);
    \node[align=left] at (2.4,-0.3) {$u_2$};
    \draw[thick] (3.6,-0.1) -- (3.6,0.1);
    \node[align=left] at (3.6,-0.3) {$u_1$};
    \node[dot, red] at (1.2,0.1) {};
    \node[dot, red] at (2.4,0.1) {};
    \node[dot, red] at (3.6,2) {};
    \draw[dashed, red] (0,0) -- (2.4,0.1) -- (3.6,2);
    \node[dot, blue] at (1.2,1) {};
    \node[dot, blue] at (2.4,2) {};
    \node[dot, blue] at (3.6,3) {};
    \draw[dashed, blue] (0,0) -- (1.2,1) -- (2.4,2) -- (3.6,3);
    \node[dot] at (1.2,0.4) {};
    \node[dot] at (2.4,1.7) {};
    \node[dot] at (3.6,1) {};
    \draw[dashed, black] (0,0) -- (1.2,0.4) -- (2.4,1.7) -- (3.6,1);
    \end{tikzpicture}
    \caption{Qualitative representation of violence frequencies, concentration levels and mean streaks as a function of $u_m$, having fixed $\underline{\eta}< \eta < \overline{\eta}$. On the $y$ axes we depict the predicted violence frequency (red) and the predicted concentration level (blue).}
    \label{fig:3ag_fights/occ_per_u}
\end{figure}
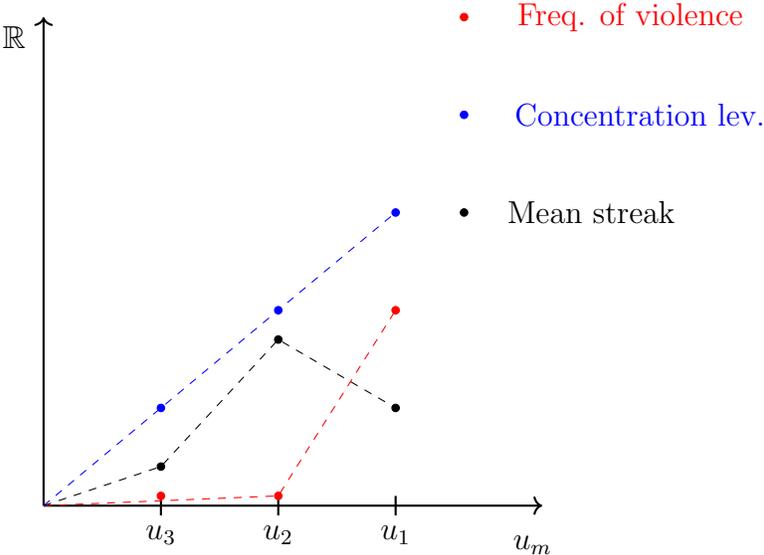

When $\eta>\overline{\eta}$, OCGs establish property rights over areas. Hence, the concentration level of each area is equal to the frequency with which OCGs leave their turf: $\frac{\eta}{1+\eta}$. When $\underline{\eta} < \eta < \overline{\eta}$, one OCG specializes in area $2$, whereas the other two occasionally clash in $1$. In such a case, the OCG that last exploited area $1$ always starts from area $1$, while the last OCG exploiting area $3$ might start from area $1$ or $3$ depending on their belief (see Lemma D.1). Hence, it must be that the concentration level of area $1$ is higher than $\frac{\eta}{1+\eta}$, the one in area $2$ is exactly $\frac{\eta}{1+\eta}$ and the one in area $3$ is lower than $\frac{\eta}{1+\eta}$. For area $3$, the concentration level has to be lower than $\frac{\eta}{1+\eta}$ since the last OCG exploiting that area occasionally goes to area $1$, and the last OCG exploiting area $1$ goes to area $3$ only when $1$ is occupied. The intuition follows for $\eta < \underline{\eta}$. In Figure~\ref{fig:3ag_fights/occ_per_u}, we qualitatively depict concentration levels for areas given $\underline{\eta} < \eta < \overline{\eta}$ to theoretically replicate those in Figure~\ref{fig:dual_axis_plot}. It is evident from comparing the two figures that $\underline{\eta} < \eta < \overline{\eta}$ seems to be an appropriate measure of criminal groups' activities in Merseyside, U.K.

\paragraph{Violence Levels.} Let \textit{violence} be the outcome of a situation in which an OCG steps into an area that has already been occupied by another OCG, and let $V_m(\eta)$ be a function that gives for each value of $\eta$ the violence frequency of area $m$.




\begin{proposition}\label{prop_3ag_fights}

$\text{ }$

\begin{itemize}
    \item If $\eta > \overline{\eta}$, $V_1(\eta)=V_3(\eta)=V_3(\eta)=0$;
    \item If $\underline{\eta} < \eta < \overline{\eta}$, $V_1(\eta) > 0$ while $V_2(\eta)=V_3(\eta)=0$;
    \item If $\eta < \underline{\eta}$, $V_1(\eta)>V_2(\eta)>0$ and $V_3(\eta)=0$.
\end{itemize}

\end{proposition}

The intuition behind the structure of violence across areas is coherent with our result on concentration levels. Indeed, when $\eta > \overline \eta$, each OCG establishes a property right over one area; hence, none of the areas exhibit violence. As the frequency of activity of criminal groups decreases, the likelihood of property rights emerging decreases, particularly in the high-revenue area. On the other side of the spectrum, when $\eta < \overline \eta$  expectations of finding OCGs out of their turf are low. Hence, any given OCG anticipates a lower probability of encountering a rival in area $1$. This, in turn, will make area $1$ more attractive. Hence, the probability of two OCGs simultaneously stepping into area $1$ is high, and so is the probability of observing violence.

An important prediction of our model is that, unless $\eta < \underline{\eta}$, OCGs will still establish \textit{de facto} property rights over area $2$. For such \textit{interval} of values of $\eta$, if an OCG always exploits area $2$, no other OCG will begin its exploration from such area, since the expected benefit of finding the area unoccupied does not compensate for the expected cost of finding it occupied. This is a very different situation than for area $1$, where the high stakes make it worth risking a conflict. Last, if  $\eta < \underline{\eta}$, all OCGs increasingly prioritize exploiting area $1$, and if $1$ is found occupied, they turn to area $2$ instead. As a result, for $\eta < \underline{\eta}$, violence also rises in area $2$.

We qualitatively depict violence frequencies in Figure~\ref{fig:3ag_fights/occ_per_u} for $\underline{\eta} < \eta < \overline{\eta}$. 
Our model predicts that as the probability of meeting rivals decreases, OCGs will increasingly target the most valuable territories, thus mirroring the empirical findings of Figure~\ref{fig:dual_axis_plot}, where we noted that the level of violence increases with the revenue of an area.  On one hand,  the model grounds the empirical observation that violence is concentrated in the high-revenue areas; on the other hand, it also suggests that sporadic violence, as recorded in middle-ranked ones, can be attributed to OCGs being able to establish property rights there.

\paragraph{Streak Length.} Last, let us now focus on the third key dimension of the empirical analysis, that is, the mean streak length.  Such a variable captures whether an OCG returns to that area once the same OCG has already exploited the area. Let $R_m(\eta)$ be a function that gives for each value of $\eta$, the expected mean streak in area $m$.
\begin{proposition}\label{prop_3ag_streak}
$\text{ }$
    \begin{itemize}
        \item If $\eta > \overline{\eta}$, $R_1(\eta)=R_2(\eta)=R_3(\eta)$;
        \item If $\underline{\eta} < \eta < \overline{\eta}$, $R_2(\eta)>R_1(\eta) > R_3(\eta)$;
        \item If $\eta < \underline{\eta}$, $R_1(\eta)>R_2(\eta) \geq R_3(\eta)$.
    \end{itemize}
\end{proposition}

Intuitively, when $\eta$ is large enough, OCGs establish \textit{de facto} rights over all three areas, and thus, each OCG always returns to the same area. This result follows from observing that when the probability of OCGs moving out of their turf is high enough, the probability of an overlap between two OCGs (hence, the cost for a collision) is so high that OCGs find it more appealing to always exploit the same area. As a result, under this configuration, streaks are equivalent across all areas.

More interestingly, for $\eta$  between  $\underline{\eta}$ and $\overline{\eta}$, OCGs manage to establish property rights only over the middle-ranked area, and occasionally collide in the higher-ranked area. Due to this outcome, the streak observed in the high-revenue area is lower than the one observed in the middle-ranked one. Moreover, the last OCG to exploit area $1$ always returns to that area. In contrast, the last OCG to exploit area $3$ might occasionally start from area $1$ instead. As a result, the streak in area $1$ must be higher than the one in area $3$.

Last, when $\eta$ is low enough (i.e.  $\eta < \underline{\eta}$), the probability that the last OCG exploiting area $1$ goes back to that area is one, while the same cannot be said for the other two areas. Therefore, the streak observed in the high-revenue area must be larger than the ones observed in the other areas. 
In Figure~\ref{fig:3ag_fights/occ_per_u} we qualitatively depict the results of Proposition~\ref{prop_3ag_streak} for a configuration such that $\eta$ picks an intermediate value. We note that the peculiar bell shape, common to both Figure \ref{fig:dual_axis_plot} and \ref{fig:3ag_fights/occ_per_u}, suggests that the complex control patterns observed in Merseyside may naturally emerge from a situation where OCGs manage to secure property rights only on a segment of areas, coinciding with the mid-revenue areas.

\paragraph{Comparative Statics.} By studying the concentration level $O_m(\cdot)$ and violence $V_m(\cdot)$ for limit values of $\eta$, we observe discontinuities which our model can rationalize. We first analyze the shape of the concentration levels.

\begin{corollary}\label{cor_3ag_occ_lim}

Consider $O_1(\eta)$, and $O_3(\eta)$.
\begin{equation*}
    \lim_{\eta \to \underline{\eta}^-} O_1(\eta) > \lim_{\eta \to \underline{\eta}^+} O_1(\eta), \text{ and } \lim_{\eta \to \underline{\eta}^-} O_3(\eta) < \lim_{\eta \to \underline{\eta}^+} O_3(\eta)
\end{equation*}
\begin{equation*}
    \lim_{\eta \to \overline{\eta}^-} O_1(\eta) > \lim_{\eta \to \overline{\eta}^+} O_1(\eta), \text{ and } \lim_{\eta \to \overline{\eta}^-} O_3(\eta) < \lim_{\eta \to \overline{\eta}^+} O_3(\eta)
\end{equation*}
\end{corollary}

 This result is a consequence of Proposition~\ref{prop_3ag_occ} and implies that as the activity level of criminal groups ($\eta$) decreases, the concentration levels of the high-revenue area increase, while the concentration levels of the low-revenue one decrease. We now analyze the violence levels.

\begin{corollary}\label{cor_3ag_fights_lim}

Consider $V_1(\eta)$, and $V_2(\eta)$.
\begin{equation*}
    \lim_{\eta \to \overline{\eta}^-} V_1(\eta) > \lim_{\eta \to \overline{\eta}^+} V_1(\eta)
\end{equation*}
\begin{equation*}
    \lim_{\eta \to \underline{\eta}^-} V_1(\eta) > \lim_{\eta \to \underline{\eta}^+} V_1(\eta), \text{ and } \lim_{\eta \to \underline{\eta}^-} V_2(\eta) > \lim_{\eta \to \underline{\eta}^+} V_2(\eta)
\end{equation*}

\end{corollary}

Similarly to the previous statement, the intuition follows from Proposition~\ref{prop_3ag_fights}. Corollary~\ref{cor_3ag_fights_lim} implies that as the criminal groups' activity ($\eta$) decreases, violence over the high-revenue area increases. This result follows from the fact that, if $\eta < \overline{\eta}$, OCGs find it more and more desirable to aim for area $1$, hence, they increase the probability of colliding in that area. Similarly for area $2$, when $\eta < \underline{\eta}$, then OCGs occasionally fight over that area, while for $\eta > \underline{\eta}$ OCGs always establish property rights over it. Hence, a decrease in OCGs' activity is not always beneficial for all the areas. We further debate the implications of these results in the discussion.

\subsection{Simulations with many areas and OCGs}\label{subsec:sim}

\begin{figure}[ht]
\centering
\includegraphics[width=.8\linewidth]{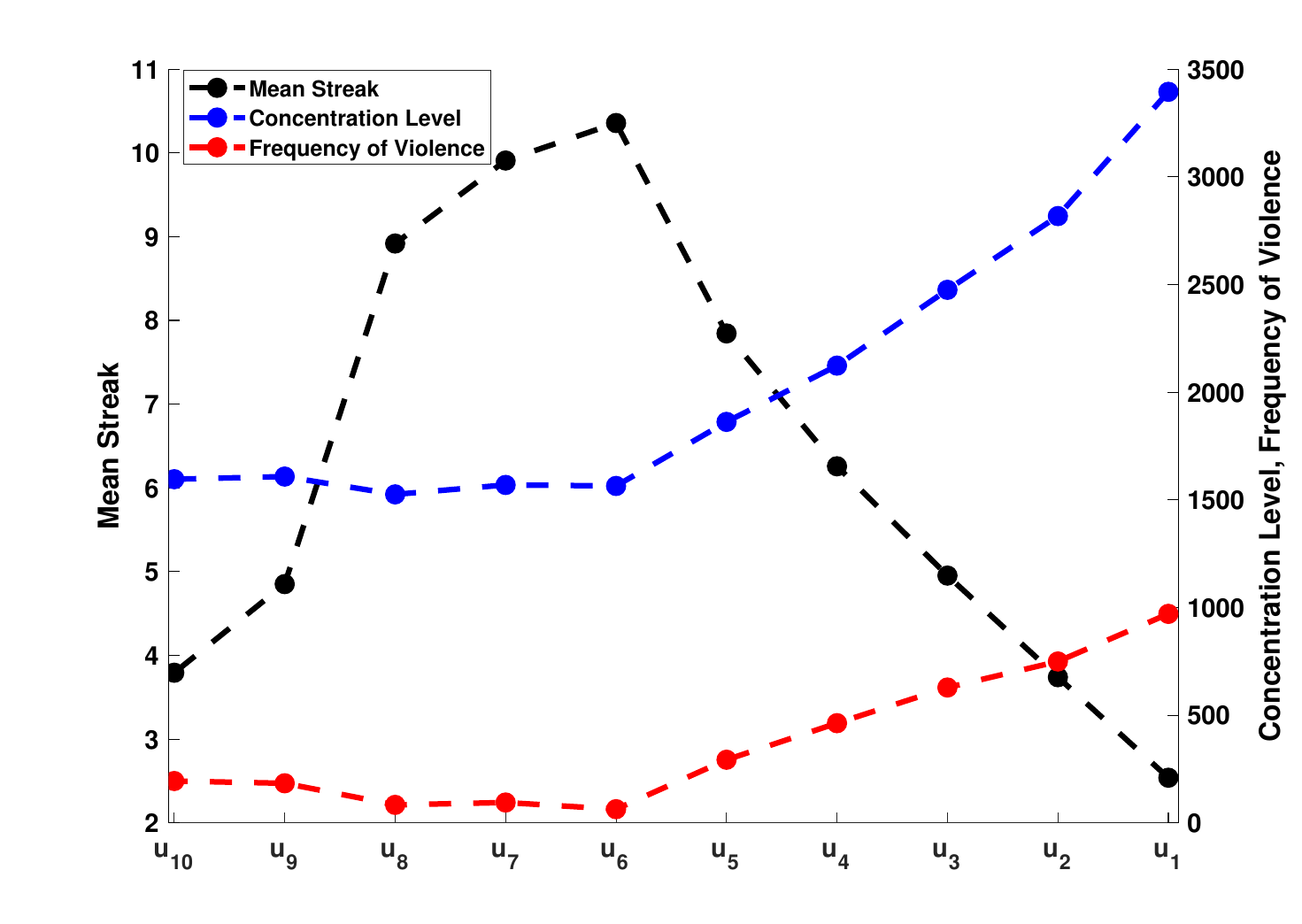}
\caption{This figure shows, for each area, the average frequency of violence,  the average OCG concentration level, and the average streak per period as obtained from the simulation of the model with setup given by $N=10$ OCGs, $M=10$ areas,  $\eta = 10$ and $T=80,000$.}
\label{fig:fights/fights_10_aggregated_fig3}
\end{figure}

Having established the intertwined effect of $\eta$ on the criminal landscape in terms of OCG concentration, violence, and streak length for a simple city made of three areas and three OCGs, we expand the model to use our theory to motivate the regularities observed in data for a setting which more closely resemble the one of Figure \ref{fig:dual_axis_plot}. We do so by generating a city made of $M=10$ areas (to mimic the grouping structure used in Figure \ref{fig:dual_axis_plot}), and explore in simulations the effect of changes to $\eta$ on the above dimensions.\footnote{A commented script containing the code and procedures used in the simulations generating Figure \ref{fig:fights/fights_10_aggregated_fig3} can be found at: \href{https://github.com/andrea-giovannetti/PropertyRightsOCGs}{https://github.com/andrea-giovannetti/PropertyRightsOCGs}.}

Let $U$ be the set of areas' values; in our simulations, for areas $1, ...10$ we impose the following (descending) value structure,  $U=\{173; 125; 100; 76; 63; 51; 42; 35; 29; 26\}$. To prevent overlaps between the thresholds theoretically identified in the previous section, we choose values of $u_j$ and $u_{j+1}$ such that $u_j - u_{j+1}$ differ for all areas. 
We populate the city with $M=10$ OCGs. Last, we fix $c=5$ across simulations, thus leaving out as manipulation variable $\eta$ only. While simulations have been robustly performed throughout a large parametric space given by $\eta \in \{0.00,0.01, 0.02, \dots,35.00\}$ (full results available in the supplementary material), results in this section are obtained by selecting $\eta^* = 10$, a value for which,  given the numerical configuration chosen above, $\underline{\eta} < \eta^* < \overline{\eta}$.  We depict the results of the simulations for $\eta = 10$ in Figure~\ref{fig:fights/fights_10_aggregated_fig3}. Note that, for $\frac{u_1 - u_10}{c} \approx 30$ and $\frac{u_9 - u_{10}}{c} \approx 0.6$, choosing $\eta=10$ ensures an intermediate regime where property rights emerge only over middle-ranked areas (see Lemmas D.2 and D.3).


Coherently with Figure~\ref{fig:dual_axis_plot} and~\ref{fig:3ag_fights/occ_per_u}, Figure~\ref{fig:fights/fights_10_aggregated_fig3} displays low levels of occupation and violence for low- and medium-ranked areas but higher values of occupation and violence for higher ranked areas. Moreover, as predicted by our theoretical model, the average number of streaks per area is low at the extremes (i.e., low and high revenue) but high in the middle. This pattern confirms our previous result that when the frequency of activity of OCGs reduces, groups establish property rights over middle-ranked areas while they engage in violent collisions in higher-ranked areas.

\section{Discussion}\label{sec:disc}

The importance of property rights goes well beyond legal institutions. Our model allows us to study the establishment of such property rights in the organized crime space by modeling such space as a dynamic environment where OCGs make choices to maximize the exploitation of areas in which drug markets are active (hence maximizing their profits). Importantly, our results allow us to study the movement of OCG-related criminal activities across different areas of a city and the associated violence dynamics across these areas \citep{bazzi2022promise}. We find that, while  OCGs aspire to establish property rights on territories, they achieve it primarily in low- and medium-revenue areas. The most remunerative areas see higher concentrations of OCGs and higher levels of violence. OCGs can have a violence-pacifier effect when they manage to establish property rights, thus showing all the properties of shadow institutions.

This result poses some major challenges to policymakers. Criminal property rights are a social bad that undermines the legitimacy, efficiency and effectiveness of legal institutions. Our model shows that, once established, they are difficult to tackle, not least because of the negative externalities they produce. Indeed, our results predict reducing the criminal activity of OCGs will increase violence and occupation levels in the areas that are considered the most profitable by criminal groups \citep[coherently with][]{moncada2013business,calderon2015beheading,massenkoff2022activity,velasco2023unintended}. Suppose that a policy maker could control the frequency $\eta$ with which OCGs go out of their turf to try to exploit the other areas of a city. The policy maker could, for example, increase the police activity across the city (or in particular areas), and hence, make it less appealing for OCGs to go out of their turf. In such a case, we can rely on the results in Corollary~\ref{cor_3ag_occ_lim} and~\ref{cor_3ag_fights_lim} to predict OCGs behavior.

If $\eta > \overline{\eta}$, the violence levels are expected to be considerably low, but the occupation level high for all areas. Thus, reducing the value of $\eta$ might be beneficial as long as $\eta > \overline{\eta}$. When $\eta < \overline{\eta}$, not only the occupation levels in the high-revenue areas but also the violence registered in such areas. Hence, a strong reduction in criminal activity could result in negative externalities for the people living in the high-revenue areas. Even more importantly, when starting from a situation of $\underline{\eta} < \eta < \overline{\eta}$, if $\eta$ reduces within that range, the effects are beneficial for all areas, but if $\eta$ reduces any further below $\underline{\eta}$, we observe negative side effects for both the high-revenue areas and the middle-ranked areas in terms of revenue. Indeed, as our results predict, when the criminal activity is sufficiently infrequent, OCGs establish no property rights over areas, meaning that violence levels expand to the middle-ranked areas in terms of revenue.


To conclude, our model, predicts that the negative externalities from reducing $\eta$ are worse when starting from a situation in which OCGs activity is not so frequent (i.e., $\underline{\eta} < \eta < \overline{\eta}$). Indeed, in such a case, the increase in violence levels spreads from the high-revenue areas to the middle-ranked ones. Our empirical results clearly fit with a parametrization of our model such that $\eta$ assumes such values. This means that when devising - and delivering interventions - there is a strong need to consider effects beyond the individual OCG and area targeted, particularly the negative side effects that could hit both the high-revenue areas and middle-ranked ones, including unintended increases in episodes of violence in such areas. As shadow institutions, OCGs can have a stabilizing effect on violence when successful in establishing property rights over illegal markets and territories; targeting them can generate destabilizing dynamics.  Therefore, the role of property rights needs to be properly grasped and integrated into (necessary) interventions to tackle organized crime even in territories that are not traditionally associated with Mafia-like organizations.

While our analysis offers novel insights into the territorial dynamics of organized crime, it is important to stress that our empirical findings are descriptive and do not establish causal relationships. Rather than identifying the effect of specific interventions or shocks, our goal is to interpret observed regularities through the lens of a dynamic model of strategic behavior. The qualitative alignment between model predictions and data patterns suggests that key features of OCG activity -- such as violence concentration and unstable control over high-revenue areas—can be understood as endogenous outcomes of opportunistic interactions under resource competition. This modeling approach complements existing empirical studies by highlighting underlying mechanisms and generating testable hypotheses for future causal analysis.

\setlength{\bibhang}{0pt}
\bibliographystyle{apalike}

\end{document}